\documentclass[prb,twocolumn,showpacs,notitlepage,amsmath,amssymb,superscriptaddress]{revtex4-2}
\usepackage{epsfig,epstopdf} 
\usepackage{graphicx}
\usepackage{dcolumn}
\usepackage{bm}
\usepackage{bbm}
\usepackage{ulem}
\usepackage{xcolor}
\usepackage{amsmath}
\usepackage{esint}
\usepackage{slashed}
\usepackage{hyperref}
\usepackage{mathrsfs}
\usepackage{subfigure}
\usepackage{verbatim}
\usepackage{dsfont}
\usepackage{float}

\begin{document}

\bibliographystyle{apsrev4-2}

\title{Bulk-Vortex Correspondence of Higher-Order Topological Superconductors}
\author{Rui-Xing Zhang}
\email{ruixing@utk.edu}
\affiliation{Department of Physics and Astronomy, University of Tennessee, Knoxville, Tennessee 37996, USA}
\affiliation{Department of Materials Science and Engineering, University of Tennessee, Knoxville, Tennessee 37996, USA}
\affiliation{Institute for Advanced Materials and Manufacturing, University of Tennessee, Knoxville, Tennessee 37920, USA}

\begin{abstract}
	Vortices in chiral topological superconductors are known for trapping Majorana zero modes as a signature response to their bulk-state topologies. In this work, we establish a similar {\it bulk-vortex correspondence} for two-dimensional $C_n$-protected class D higher-order topological superconductors, in which a quantum vortex will universally trap a pair of $C_n$-protected Majorana bound states with $\hat{z}$-directional angular momenta $J_z=0$ and $J_z=n/2$, respectively. This intriguing one-to-one mapping between anomalous vortex modes and higher-order topology can be systematically derived through a model-independent patch construction approach. As a proof of concept, we present an exactly solvable model to confirm the proposed vortex Majorana modes. Our theory establishes vortices as an unprecedented experimental measure of higher-order topology in superconductors.
\end{abstract}

\maketitle

{\it Introduction} - The adventure of topological superconductors (TSCs) started from a remarkable theoretical prediction, that edges and vortices of a two-dimensional (2D) chiral superconductor can harbor 1D chiral Majorana modes and 0D Majorana zero modes (MZMs), respectively~\cite{read2000chiral,ivanov2001nonabelian,kitaev2001unpaired,nayak2008anyon,jay2010chiral,jay2010nonabelian,sato2009chiral,qi2010chiral,alicea2012new}. This necessary bondings between bulk-state topology and edge/vortex Majorana modes are dubbed the bulk-boundary correspondence (BBC) and bulk-vortex correspondence (BVC). In particular, BBC of chiral TSCs originates from a nonvanishing Chern number for the Bogoliubov-de Gennes (BdG) Hamiltonian~\cite{niu1985quantized}, thus yielding intuitive understandings such as Laughlin's argument~\cite{laughlin1981pumping} or evolutions of Wannier function centers~\cite{thouless1983quantization,vanderbilt1993polarization,vanderbilt2006wannier,fu2006time}. On the other hand, a vortex can be effectively viewed as a circular-shaped inner boundary with a radius of the superconducting coherence length~\cite{alicea2012new}. Then a vortex MZM is essentially a Majorana edge state with a special angular momentum. This implies an equivalence between BBC and BVC for chiral TSCs. 

Since then, enormous research effort has been devoted to exploring other topological possibilities of superconductivity, which has led to a plethora of exciting ideas such as time-reversal-invariant TSCs~\cite{qi2009time,qi2010topological,zhang2013time,zhang2021intrinsic}, mirror TSCs~\cite{zhang2013topo,langbehn2017reflection}, Dirac/Weyl TSCs~\cite{meng2012weyl,yang2014dirac,kobayashi2015dirac}, and recently higher-order TSCs~\cite{teo2013existence,benalcazar2014class,wang2018weak,shapourian2018topo,wang2018high,yan2018majorana,you2018higher,zhang2019helical,ghorashi2019second,zhang2019higher,volpez2019second,liu2018majorana,yan2019higher,hsu2020inversion,pan2019lattice,zhang2020kitaev,zhang2022strongly,hu2022higher}, etc. Most studies, however, focus on the BBC of TSCs~\cite{khalaf2018higher,trifunovic2019higher,titus2020atomic,roberts2020second}, while little is known about their BVC aspects. Meanwhile, we note that some TSCs may not possess a BVC at all, since a vortex necessarily breaks the protecting symmetry of the bulk topology, e.g., helical TSCs and time-reversal symmetry~\cite{wu2017majorana}. Therefore, it has remained an interesting open question that which class of {\it non-chiral} TSCs can feature a BVC, as well as bulk-topology-enforced vortex Majorana modes.

\begin{figure}[t]
	\centering
	\includegraphics[width=\linewidth]{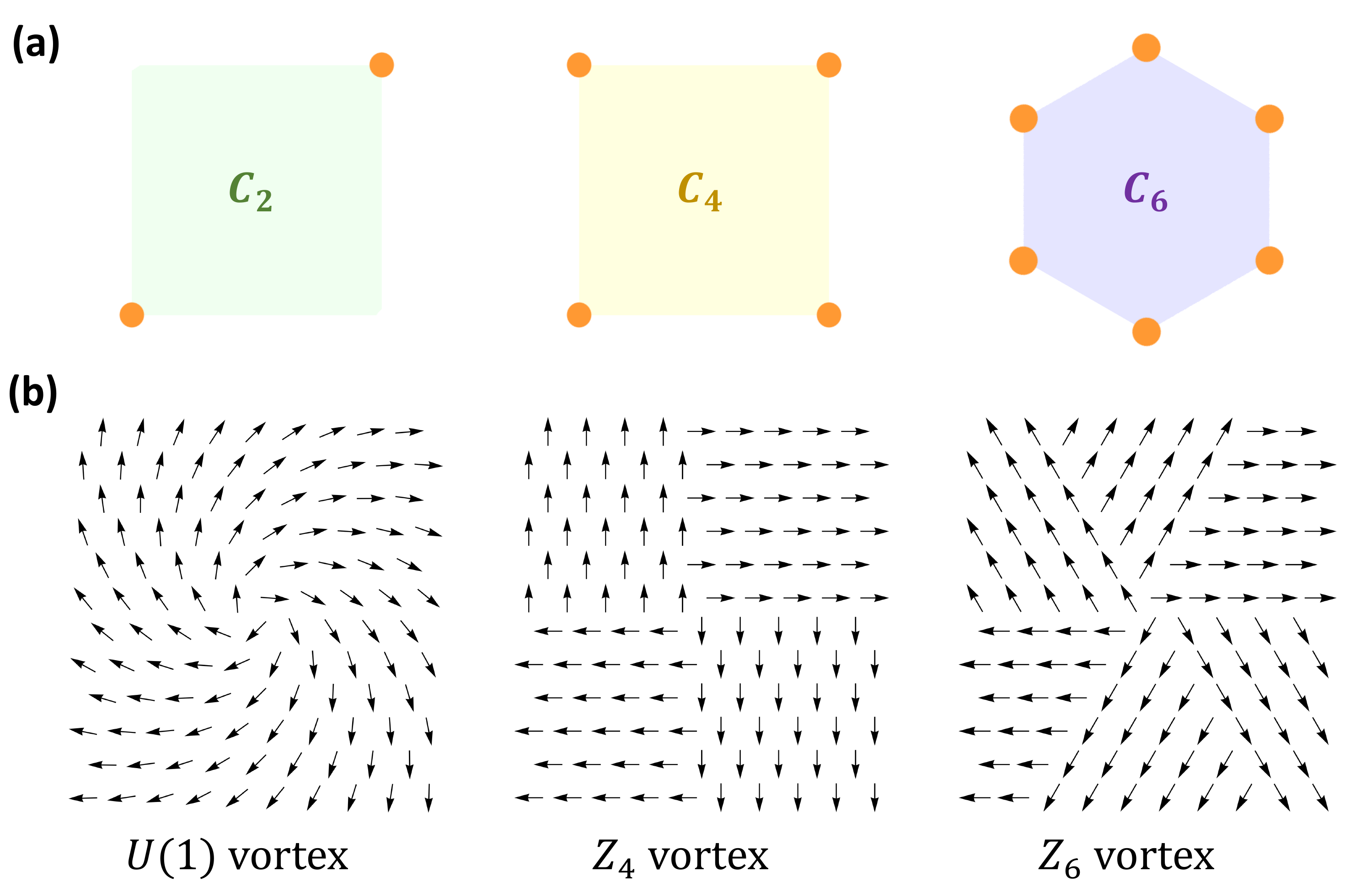}
	\caption{(a) Schematics of HOTSC$_n$ respecting an $n$-fold rotation symmetry $C_n$. Orange circles denote corner-localized MZMs. (b) A continous $U(1)$ vortex can be adiabatically deformed into a discrete $\mathbb{Z}_4$ or $\mathbb{Z}_6$ vortex, while respecting both the lattice symmetries and the bulk energy gap. The arrow shows the orientation of the local superconducting phase.}	
	\label{fig_vortex}
\end{figure}

Our main finding in this work is that a similar BVC generally exists for 2D class D higher-order topological superconductors (HOTSCs). These HOTSCs are essentially $C_n$-protected crystalline TSCs that feature a 2nd-order BBC~\cite{teo2013existence,benalcazar2014class}, which manifests in the presence of $n$ Majorana modes bound to the system's geometric corners. Remarkably, our BVC dictates that every superconducting vortex will contribute a $C_n$-protected pair of vortex MZMs to the system, in addition to the corner MZMs. We provide a patch construction approach to derive the BVC and further find the two vortex MZMs to carry angular momenta $J_z=0$ and $J_z=n/2$, respectively. Crucially, the patch construction offers an explicit mapping between corner Majorana degrees of freedom and those inside the vortex, further establishing the vortex MZMs as a direct consequence of bulk higher-order topology. We further provide a minimal model of HOTSC to analytically demonstrate the BVC. 

\begin{figure*}[t]
	\centering
	\includegraphics[width=\linewidth]{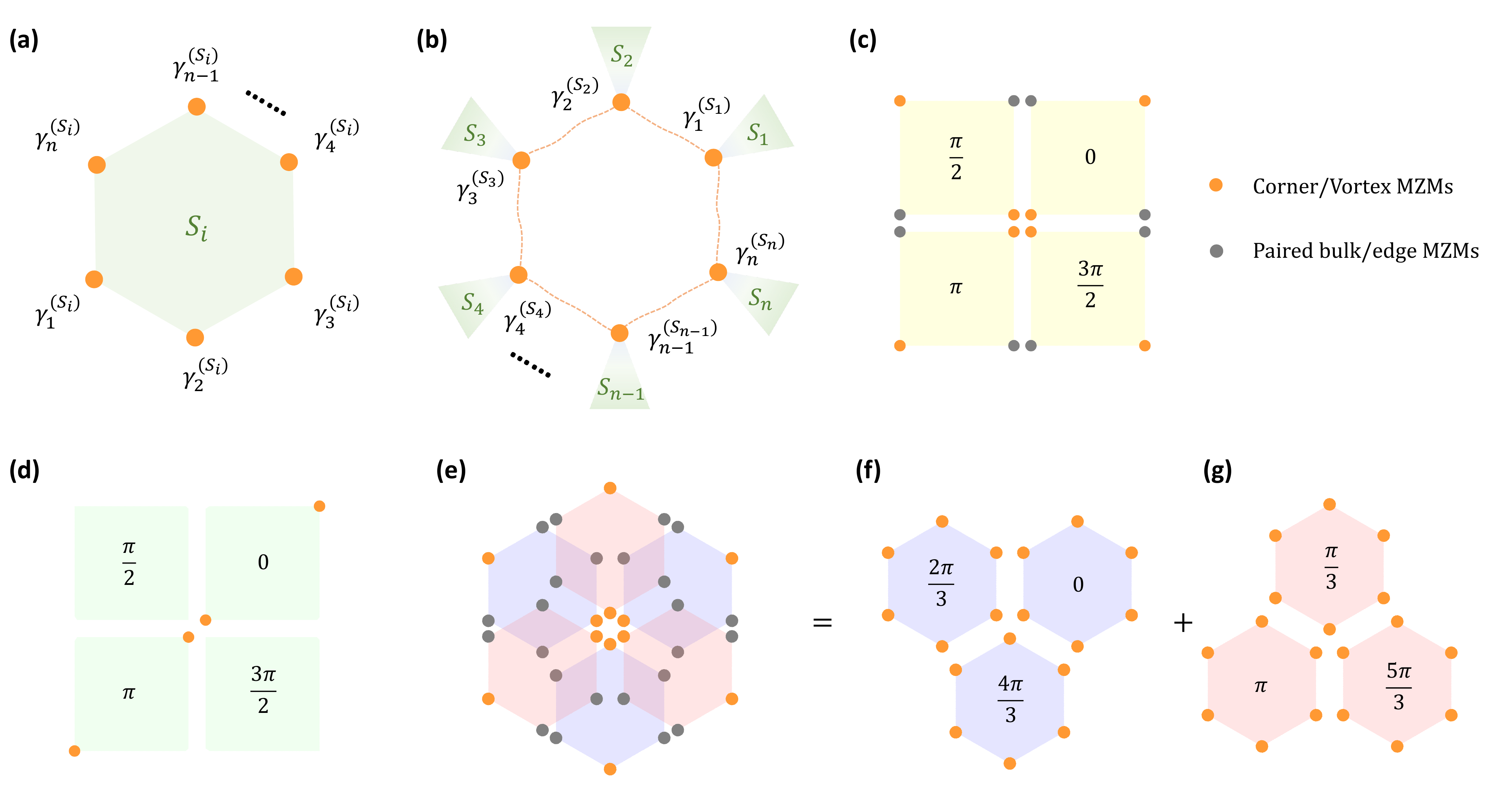}
	\caption{(a) An HOTSC$_n$ patch $S_i$ with its Majorana-carrying corners. (b) A schematic of $S_\Sigma$ that consists of $n$ small patches. Note that the $n$-Majorana ring is formed by the collection of the $i$th corner MZM of the patch $S_i$, i.e., $\gamma_i^{(S_i)}$. (c) and (d) show how a $\mathbb{Z}_4$ vortex can be implemented in the patch construction for HOTSC$_4$ and HOTSC$_2$, respectively. (e) shows a $\mathbb{Z}_6$ vortex of HOTSC$_6$ with a 6-patch geometry, which involves a pair of twisted $\mathbb{Z}_3$ vortices shown in (f) and (g). The black circles are corner MZMs from smaller patches that can be trivialized while combining with other patches. Orange circles represent either unpaired corner MZMs or vortex MZMs that form an $n$-Majorana ring.}	
	\label{fig_glue}
\end{figure*}

{\it HOTSC$_n$ \& $\mathbb{Z}_m$ Vortex} - The defining boundary signature for $C_n$-protected HOTSC (dubbed HOTSC$_n$) is a set of $n$ MZMs that are $C_n$-related. In a finite-size regular polygon geometry, the boundary MZMs are likely to reside around the polygon corners and are known as {\it corner MZMs}. The corner MZMs cannot be removed without either (i) breaking $C_n$ symmetry; or (ii) closing the bulk-state energy gap. Notably, HOTSC$_n$ generally admits a $\mathbb{Z}_2$ topological classification and the topological phase hosts an odd number of MZMs at each geometric corner. Schematics of $C_{2,4,6}$-invariant HOTSCs are shown in Fig.~\ref{fig_vortex}~(a).

Formally, an HOTSC$_n$ can be described by a Bogoliubov-de Gennes (BdG) Hamiltonian, 
\begin{equation}
	H({\bf k}) = \begin{pmatrix}
		h_0({\bf k}) & \Delta({\bf k}) \\
		\Delta({\bf k}) & -h_0(-{\bf k})^T \\
	\end{pmatrix},
   \label{eq-ham-H}
\end{equation}
where the pairing function follows $\Delta^T({\bf k}) = -\Delta(-{\bf k})$. A superconducting vortex can be introduced by updating $\Delta({\bf k})\rightarrow \Delta(r, \theta)e^{i\theta}$, where $(r, \theta)$ denote the in-plane polar coordinates. Without loss of generality, one can deform such a continous $U(1)$ vortex pattern into a ``$\mathbb{Z}_{m}$ vortex" to be congruent with the discrete lattice symmetry of HOTSCs. As shown in Fig.~\ref{fig_vortex}~(b), a $\mathbb{Z}_m$ vortex consists of $m$ patches of superconductors, with the $j$-th patch carrying a locally uniform superconducting (SC) phase factor of $\text{exp}(i \frac{2\pi }{m}j)$~\cite{fu2008vortex}. It is clear that the deformation from a $U(1)$ vortex to a $\mathbb{Z}_m$ vortex preserves the bulk-state gap structure, the vorticity, and the $C_m$ rotation symmetry. Therefore, topological aspects of both the bulk and vortex states will remain invariant under this $U(1)\rightarrow \mathbb{Z}_m$ vortex deformation. 

{\it From Corner to Vortex} - We now introduce a {\it patch construction} approach for HOTSC$_n$  that maps corner-originated Majorana degrees of freedom to those trapped by a vortex. Take $n=4$ as an example.  We start with four identical patches $S_{i}$ ($i=1,2,3,4$) of HOTSC$_4$, each of which features a square geometry with four corner MZMs. When we glue the four patches following Fig.~\ref{fig_glue}~(c), they merge into a single large patch $S_\Sigma$. Without any vortex, all previous corner MZMs of small patches will now be paired up and lose their Majorana nature, except for the four living at the corners of the large patch. $S_\Sigma$ is, therefore, an HOTSC$_4$ as well. As we discussed earlier, a $\mathbb{Z}_4$ vortex can be introduced by assigning each path $S_j$ with a SC phase of $j\pi/2$, as schematically shown in Fig.~\ref{fig_glue}~(c). The four MZMs at the center of $S_\Sigma$ now manifest as vortex-localized MZMs and we dub them as a {\it 4-Majorana ring}. 

Vortices in HOTSC$_2$ and HOTSC$_6$ can be understood in a similar fashion. For HOTSC$_2$, we again consider four patches $S_{1,2,3,4}$, with $S_{1,3}$ and $S_{2,4}$ carrying HOTSC$_2$ and trivial SC phases, respectively. Introducing a $\mathbb{Z}_4$ vortex following Fig.~\ref{fig_glue}~(d), we find that the central two MZMs will contribute to vortex physics, forming a $2$-Majorana ring. As shown in Fig.~\ref{fig_glue}~(e), the patch geometry for HOTSC$_6$ involves a pair of $3$-patches that admit a relative twist of $\frac{\pi}{3}$, to ensure the $C_6$ invariance for each individual patch. Turning on a $\mathbb{Z}_6$ vortex generates a $6$-Majorana ring at the vortex center. 

{\it Bulk-Vortex Correspondence} - We are now ready to study these $n$-Majorana rings to reveal the nature of the corresponding vortex bound states. Following Fig.~\ref{fig_glue}~(a), we label a corner MZM at the $i$-th corner of the patch $S_j$ as $\gamma_{i}^{(S_j)}$. Then the $n$-Majorana ring at each vortex is formed by $\gamma_{j}^{(S_j)}$ with $j=1,2,...,n$, as schematically shown in Fig.~\ref{fig_glue}~(b). For convenience, we will adopt the shorthand notation $\gamma_i \equiv \gamma_{i}^{(S_i)}$ throughout the main text. 

Define the Majorana ring basis as $\Gamma_n = (\gamma_1, \gamma_2, ..., \gamma_n)^T$ and the particle-hole symmetry (PHS) is $\Xi = \mathbbm{1}_{n} {\cal K}$ with ${\cal K}$ being the complex conjugation. In the Supplemental Material, we prove that MZMs from different patches are related to one another via
\begin{equation}
	C_n \gamma_{i}^{(S_j)} C_n^{-1} = \gamma_{i+1}^{(S_{j+1})},
	\label{eq-Cn-MZM-vortex}
\end{equation}
when the SC vortex is present. In particular, we have $(C_n)^n = -(-1)^v$ with $v$ being the vorticity of the SC vortex. The matrix representation of $C_n$ symmetry is
\begin{equation}
	R_n = \begin{pmatrix}
		 0& 1 & 0 & \dots & 0 \\
		 0 & 0 & 1 &  \dots & 0 \\
		 \vdots & \vdots & \vdots &  \ddots & \vdots \\
		 0 & 0 & 0 &  \dots & 1 \\
		 1 & 0 & 0 &  \dots & 0 \\
	\end{pmatrix}_{n\times n}, \ \ \text{with }
(R_n)^n = \mathbbm{1}_{n\times n}.
\end{equation}
The most general single-particle Hamiltonian for the $n$-Majorana ring is
${\cal H}_n = i \sum_{j<k} t_{j,k} \gamma_j \gamma_k$, which contains $\binom{n}{2}$ Majorana pairing terms. Since $C_n$ symmetry requires $t_{j, j+m} =  t_{j', j'+m} \equiv t_{\frac{n}{m}}$, ${\cal H}_n$ can thus be reduced to
\begin{equation}
	{\cal H}_n = \sum_{m=1}^{n/2} t_{\frac{n}{m}} h_{n,m},\ \ \text{with  }\ h_{n,m}=i \sum_{j=1}^{n}  \gamma_{j} \gamma_{j+m}.
\end{equation}
Notice that $h_{n,m}$ couples two MZMs that are $(C_n)^m$ related, which leads to an interesting identity that
\begin{equation}
	h_{n,m} = i\Gamma_n^{\dagger} [(R_n)^m - (R_n^{\dagger})^m] \Gamma_n.
\end{equation}
Given that $C_2 = C_2^{-1}$, we immediately arrive at $h_{n,\frac{n}{2}} = 0$. Therefore, we have
\begin{eqnarray}
	&& H_2 = 0,\nonumber \\
	&& H_4 = it_4 [R_4 - R_4^{-1}],  \nonumber \\
	&& H_6 = i t_6 [R_6 - R_6^{-1}] + i t_3 [R_6^2 - (R_6^2)^{-1}].  \label{eq-Hn}
\end{eqnarray}
where $H_n$ is the matrix representation of ${\cal H}_n$ under $\Gamma_n$.

Eq.~\ref{eq-Cn-MZM-vortex} further implies that the $n$-Majorana ring forms a regular representation of $C_n$ group, so that the MZMs can be further superposed to form a complete set of $C_n$ eigenstates $\{|J_z\rangle\}$ that covers all 1D irreps of this $n$-fold cyclic group, where $J_z\in\{0,1,...,n-1\}$ is the $\hat{z}$-component angular momentum. The energy eigen-equation now becomes $H_n|J_z\rangle = E_{n, J_z} |J_z\rangle$ with,
\begin{eqnarray}
	&& E_{2,J_z} = 0,\ \ \ \ \ E_{4, J_z} = -2 t_4 \sin (\frac{\pi}{2} J_z),\nonumber \\
	&& E_{6,J_z} = -2 t_6 \sin (\frac{\pi}{3} J_z) -2 t_3 \sin (\frac{2\pi}{3} J_z),
\end{eqnarray}
leading to two {\it zero-energy} Majorana solutions: $|J_z=0\rangle$ and $|J_z=\frac{n}{2}\rangle$. From a symmetry perspective, $|J_z=0,\frac{n}{2}\rangle$ belong to the only two PHS-invariant $C_n$ irreps~\cite{liu2014lattice,hu2022topological}, a basic requirement for Majorana wavefunctions. We thus conclude a {\it bulk-vortex correspondence}: 
\begin{itemize}
	\item A 2D $C_n$-protected class D HOTSC hosts a pair of vortex MZMs with $J_z=0$ and $J_z=\frac{n}{2}$, respectively. 
\end{itemize}

{\it Minimal Model} - As a proof of concept, we now provide a minimal model of HOTSC$_n$ that yields analytical solutions to demonstrate its BVC. The ${\bf k}$-space BdG Hamiltonian of this four-band model is
\begin{equation}
	H_\text{BL}({\bf k}) = \begin{pmatrix}
		h_+ ({\bf k}) & h_c({\bf k}) \\
		h_c^{\dagger}({\bf k}) & h_- ({\bf k})
	\end{pmatrix}.
\end{equation}
where $h_{\pm}({\bf k}) = \Delta (k_x \tau_x + k_y \tau_y) \pm M({\bf k}) \tau_z$. Here $M({\bf k}) = m(k_x^2+k_y^2)-\mu$ and $\mu$ is the chemical potential. Therefore, $H_\text{BL}({\bf k})$ manifests as a ``bilayer" superconductor with a layer-dependent BdG Chern number ${\cal C}_{\pm}=\mp\text{sgn}(m)$ when $m\mu>0$~\cite{zhang2020kitaev}. Note that $h_{\pm}({\bf k})$ will contribute to a pair of counterpropagating chiral Majorana edge modes, which will be generally spoiled by introducing $h_c({\bf k})$. As we will show next, under certain circumstances, $C_n$ enforces the inability of $h_c({\bf k})$ to achieve a thorough edge gap, and the residue Majorana degrees of freedom further become corner MZMs. 

The ``interlayer" coupling $h_c({\bf k})$ explicitly depends on the Nambu basis $\Psi_{\pm}$ for $h_{\pm}$. We express $\Psi_{\pm}$ in the angular momentum basis $|J_z\rangle$ and find that $C_n$ requires $\Psi_{\pm}=(|J_z\rangle, |J_z-1\rangle)^T$ with $J_z$ being half-integer valued. In addition, $\Psi_{\pm}$ should be PHS-invariant and this leads to $-J_z \equiv J_z-1$ (mod $n$). It is easy to show that there exist two inequivalent basis solutions for $\Psi_{\pm}$,
\begin{eqnarray}
	\Psi_\nu = (|-\nu+\frac{1}{2}\rangle, |\nu-\frac{1}{2}\rangle)^T,\ \ \text{for }\nu\in\{0,\frac{n}{2}\}.
\end{eqnarray}
Therefore, $\Psi_\text{BL} = (\Psi_{\nu}, \Psi_{\nu'})^T$ with $\nu,\nu'\in\{0,\frac{n}{2}\}$ features four basis choices that are compactible with both PHS and $C_n$. In the Supplemental Material, we analytically solve the edge theory in the polar coordinates and find that an HOTSC$_n$ phase only shows up when $\nu + \nu'=\frac{n}{2}$. In particular, for $(\nu,\nu')=(0,\frac{n}{2})$, we find the coupling matrix $h_c({\bf k})$ to be
\begin{equation}
	h_c({\bf k}) = \begin{pmatrix}
		\alpha k_-^{\frac{n}{2}} & \beta k_+^{\frac{n}{2}-1} \\
		(-1)^{\frac{n}{2}}\beta k_-^{\frac{n}{2}-1} & (-1)^{(\frac{n}{2}+1)}\alpha k_+^{\frac{n}{2}} \\
	\end{pmatrix},
\end{equation}
where $k_{\pm}=k_x\pm i k_y$. In the polar coordinates $(r,\theta)$, $h_c(r,\theta)$ is found to generate an edge mass term proportional to $\sin n\theta/2$ (for $n=2,6$) or $\cos n\theta/2$ (for $n=4$). As a result, the edge gap closes whenever the polar angle $\theta = 2\pi l/n$ (for $n=2,6$) or $\theta=(2l+1)\pi/n$ (for $n=4$), with $l=0,1,2,...,n-1$. This corresponds to $n$ corner MZMs, a defining boundary characteristics of HOTSC$_n$. 

The vortex bound state of $H_\text{BL}$ can be similarly solved by treating $h_c({\bf k})$ perturbatively. In the decoupling limit with $h_c({\bf k})=0$, the BVC of chiral TSCs dictates the appearance of two vortex MZMs contributed by $h_{\pm}$, respectively~\cite{read2000chiral}. The vortex zero-mode wavefunction can be solved by decorating $h_{\pm}({\bf k})$ with a superconducting phase winding. For example, the previous uniform chiral $p$-wave pairing $\Delta k_-$ shall now be updated to $\{\Delta(r)e^{i\theta}, \partial_{\bar{z}}\}$, where $\bar{z} = x-iy$ and $\Delta(r) = \Delta_0 \tanh(r/\xi)$. Here $\xi$ is the superconducting coherence length and it decides the size of a quantum vortex. It is staightfowrad to show that vortex zero-mode solutions for $h_{\pm}$ is $|\gamma_+(r,\theta)\rangle = |\gamma_-(r,\theta)\rangle = F(r) [e^{i\pi/4}, e^{-i\pi/4}]^T$~\cite{kopnin1991mutual,volovik1999fermion,sumanta2007index,gurarie2007zero,sumanta2007quantum,cheng2009splitting}, where $F(r)$ describes the spatial feature of the Majorana wavefunction. Note that the equality between $|\gamma_+\rangle$ and $|\gamma_-\rangle$ arises from $\Xi h_{+}({\bf k}) \Xi^{-1} = h_-({\bf k})$. The expression of the zero-mode wavefunctions immediately implies that vortex MZMs $\gamma_{\pm}$ will inherit the angular momenta $J_z$ from $\Psi_\nu$, with an additional half-integer shift due to the vortex phase winding. For the trivial SC phase with $\nu=\nu'=0$ or $\frac{n}{2}$, the two vortex MZMs carry the same $J_z\in\{0,\frac{n}{2}\}$, which will immediately get hybridized and lose their Majorana nature once $h_c$ is turned on. On the contrary, for the HOTSC$_n$ phase with $(\nu,\nu')=(0,\frac{\pi}{2})$, we find that
\begin{eqnarray}
	\hat{J}_z |\gamma_+\rangle = 0,\ \ \hat{J_z}|\gamma_-\rangle = \frac{n}{2} |\gamma_-\rangle.
\end{eqnarray} 
Therefore, $\gamma_{\pm}$ are isolated from each other, thanks to their distinct $C_n$ indices, and this symmetry protection persists even in the presence of $h_c$. This exactly agrees with our prediction of BVC for HOTSC$_n$.

{\it Discussions} - To summarize, we derive a general bulk-vortex correspondence for 2D $C_n$-protected HOTSCs in symmetry class D through a patch construction approach. In particular, the BVC concretely predicts both the number and the angular momentum information of MZMs trapped by a superconducting vortex, which is analytically confirmed by solving a minimal HOTSC$_n$ model. We highlight that our derivation of BVC does not depend on any model details, and should hold for general higher-order topological systems that fall into our target symmetry classes. 

In practice, experimental detection of crystalline topological phases is generally challenging, since their gapless boundary modes are usually sensitive to the local boundary conditions. For example, the topological robustness of corner MZMs in an HOTSC$_n$ is only well-defined when the sample geometry delicately preserves $C_n$. Real-world crystals, however, are often irregularly shaped, and such asymmetric boundaries could possibly spoil the expected corner MZMs. On the contrary, a superconducting vortex is usually circularly shaped and generally respects the desired crystalline symmetry. Namely, detection of vortex MZMs in an HOTSC could be much more feasible than the corner ones, with the help of state-of-the-art scanning tunneling microscopy technique. Therefore, the BVC proposed in this work paves the way for {\it locally} probing superconducting higher-order topology with vortices, even for irregular crystal samples whose corner signatures are ambiguous.  

Finally, our BVC serves as a good starting point to explore vortex physics of crystalline topological superconductors in other symmetry classes and/or in other spatial dimensions. For example, stacking 2D HOTSC$_n$ could lead to a plethora of 3D topological phases. These include weak HOTSC$_n$ with flat-band Majorana hinge modes, higher-order Weyl/Dirac superconductors~\cite{wieder2020strong,zhang2020dirac}, etc, all of which should exhibit 1D dispersing vortex Majorana modes as a derivative of our 2D BVC. Besides space group symmetries, it would also be intriguing to ask whether magnetic group symmetries~\cite{fang2014new,zhang2015magnetic,li2020pfaffian} can support similar BVC for crystalline TSC systems. We leave these interesting directions for future works.     

\begin{acknowledgements}
	 We thank J. Yu, X.-Q. Sun, and L.-H. Hu for helpful discussions. We are particularly indebted to Y. You for inspiring discussions on their unpublished results~\cite{you2022unpublished} that motivated us to consider the vortex problem of HOTSCs. This work is supported by a start-up fund at the University of Tennessee.
\end{acknowledgements}

\bibliographystyle{apsrev4-2}
\bibliography{BVC}

\begin{thebibliography}{62}%
\makeatletter
\providecommand \@ifxundefined [1]{%
 \@ifx{#1\undefined}
}%
\providecommand \@ifnum [1]{%
 \ifnum #1\expandafter \@firstoftwo
 \else \expandafter \@secondoftwo
 \fi
}%
\providecommand \@ifx [1]{%
 \ifx #1\expandafter \@firstoftwo
 \else \expandafter \@secondoftwo
 \fi
}%
\providecommand \natexlab [1]{#1}%
\providecommand \enquote  [1]{``#1''}%
\providecommand \bibnamefont  [1]{#1}%
\providecommand \bibfnamefont [1]{#1}%
\providecommand \citenamefont [1]{#1}%
\providecommand \href@noop [0]{\@secondoftwo}%
\providecommand \href [0]{\begingroup \@sanitize@url \@href}%
\providecommand \@href[1]{\@@startlink{#1}\@@href}%
\providecommand \@@href[1]{\endgroup#1\@@endlink}%
\providecommand \@sanitize@url [0]{\catcode `\\12\catcode `\$12\catcode
  `\&12\catcode `\#12\catcode `\^12\catcode `\_12\catcode `\%12\relax}%
\providecommand \@@startlink[1]{}%
\providecommand \@@endlink[0]{}%
\providecommand \url  [0]{\begingroup\@sanitize@url \@url }%
\providecommand \@url [1]{\endgroup\@href {#1}{\urlprefix }}%
\providecommand \urlprefix  [0]{URL }%
\providecommand \Eprint [0]{\href }%
\providecommand \doibase [0]{https://doi.org/}%
\providecommand \selectlanguage [0]{\@gobble}%
\providecommand \bibinfo  [0]{\@secondoftwo}%
\providecommand \bibfield  [0]{\@secondoftwo}%
\providecommand \translation [1]{[#1]}%
\providecommand \BibitemOpen [0]{}%
\providecommand \bibitemStop [0]{}%
\providecommand \bibitemNoStop [0]{.\EOS\space}%
\providecommand \EOS [0]{\spacefactor3000\relax}%
\providecommand \BibitemShut  [1]{\csname bibitem#1\endcsname}%
\let\auto@bib@innerbib\@empty
\bibitem [{\citenamefont {Read}\ and\ \citenamefont
  {Green}(2000)}]{read2000chiral}%
  \BibitemOpen
  \bibfield  {author} {\bibinfo {author} {\bibfnamefont {N.}~\bibnamefont
  {Read}}\ and\ \bibinfo {author} {\bibfnamefont {D.}~\bibnamefont {Green}},\
  }\href {https://doi.org/10.1103/PhysRevB.61.10267} {\bibfield  {journal}
  {\bibinfo  {journal} {Phys. Rev. B}\ }\textbf {\bibinfo {volume} {61}},\
  \bibinfo {pages} {10267} (\bibinfo {year} {2000})}\BibitemShut {NoStop}%
\bibitem [{\citenamefont {Ivanov}(2001)}]{ivanov2001nonabelian}%
  \BibitemOpen
  \bibfield  {author} {\bibinfo {author} {\bibfnamefont {D.~A.}\ \bibnamefont
  {Ivanov}},\ }\href {https://doi.org/10.1103/PhysRevLett.86.268} {\bibfield
  {journal} {\bibinfo  {journal} {Phys. Rev. Lett.}\ }\textbf {\bibinfo
  {volume} {86}},\ \bibinfo {pages} {268} (\bibinfo {year} {2001})}\BibitemShut
  {NoStop}%
\bibitem [{\citenamefont {Kitaev}(2001)}]{kitaev2001unpaired}%
  \BibitemOpen
  \bibfield  {author} {\bibinfo {author} {\bibfnamefont {A.~Y.}\ \bibnamefont
  {Kitaev}},\ }\href {https://doi.org/10.1070/1063-7869/44/10S/S29} {\bibfield
  {journal} {\bibinfo  {journal} {Physics-uspekhi}\ }\textbf {\bibinfo {volume}
  {44}},\ \bibinfo {pages} {131} (\bibinfo {year} {2001})}\BibitemShut
  {NoStop}%
\bibitem [{\citenamefont {Nayak}\ \emph {et~al.}(2008)\citenamefont {Nayak},
  \citenamefont {Simon}, \citenamefont {Stern}, \citenamefont {Freedman},\ and\
  \citenamefont {Das~Sarma}}]{nayak2008anyon}%
  \BibitemOpen
  \bibfield  {author} {\bibinfo {author} {\bibfnamefont {C.}~\bibnamefont
  {Nayak}}, \bibinfo {author} {\bibfnamefont {S.~H.}\ \bibnamefont {Simon}},
  \bibinfo {author} {\bibfnamefont {A.}~\bibnamefont {Stern}}, \bibinfo
  {author} {\bibfnamefont {M.}~\bibnamefont {Freedman}},\ and\ \bibinfo
  {author} {\bibfnamefont {S.}~\bibnamefont {Das~Sarma}},\ }\href
  {https://doi.org/10.1103/RevModPhys.80.1083} {\bibfield  {journal} {\bibinfo
  {journal} {Rev. Mod. Phys.}\ }\textbf {\bibinfo {volume} {80}},\ \bibinfo
  {pages} {1083} (\bibinfo {year} {2008})}\BibitemShut {NoStop}%
\bibitem [{\citenamefont {Sau}\ \emph {et~al.}(2010{\natexlab{a}})\citenamefont
  {Sau}, \citenamefont {Lutchyn}, \citenamefont {Tewari},\ and\ \citenamefont
  {Das~Sarma}}]{jay2010chiral}%
  \BibitemOpen
  \bibfield  {author} {\bibinfo {author} {\bibfnamefont {J.~D.}\ \bibnamefont
  {Sau}}, \bibinfo {author} {\bibfnamefont {R.~M.}\ \bibnamefont {Lutchyn}},
  \bibinfo {author} {\bibfnamefont {S.}~\bibnamefont {Tewari}},\ and\ \bibinfo
  {author} {\bibfnamefont {S.}~\bibnamefont {Das~Sarma}},\ }\href
  {https://doi.org/10.1103/PhysRevLett.104.040502} {\bibfield  {journal}
  {\bibinfo  {journal} {Phys. Rev. Lett.}\ }\textbf {\bibinfo {volume} {104}},\
  \bibinfo {pages} {040502} (\bibinfo {year} {2010}{\natexlab{a}})}\BibitemShut
  {NoStop}%
\bibitem [{\citenamefont {Sau}\ \emph {et~al.}(2010{\natexlab{b}})\citenamefont
  {Sau}, \citenamefont {Tewari}, \citenamefont {Lutchyn}, \citenamefont
  {Stanescu},\ and\ \citenamefont {Das~Sarma}}]{jay2010nonabelian}%
  \BibitemOpen
  \bibfield  {author} {\bibinfo {author} {\bibfnamefont {J.~D.}\ \bibnamefont
  {Sau}}, \bibinfo {author} {\bibfnamefont {S.}~\bibnamefont {Tewari}},
  \bibinfo {author} {\bibfnamefont {R.~M.}\ \bibnamefont {Lutchyn}}, \bibinfo
  {author} {\bibfnamefont {T.~D.}\ \bibnamefont {Stanescu}},\ and\ \bibinfo
  {author} {\bibfnamefont {S.}~\bibnamefont {Das~Sarma}},\ }\href
  {https://doi.org/10.1103/PhysRevB.82.214509} {\bibfield  {journal} {\bibinfo
  {journal} {Phys. Rev. B}\ }\textbf {\bibinfo {volume} {82}},\ \bibinfo
  {pages} {214509} (\bibinfo {year} {2010}{\natexlab{b}})}\BibitemShut
  {NoStop}%
\bibitem [{\citenamefont {Sato}\ \emph {et~al.}(2009)\citenamefont {Sato},
  \citenamefont {Takahashi},\ and\ \citenamefont {Fujimoto}}]{sato2009chiral}%
  \BibitemOpen
  \bibfield  {author} {\bibinfo {author} {\bibfnamefont {M.}~\bibnamefont
  {Sato}}, \bibinfo {author} {\bibfnamefont {Y.}~\bibnamefont {Takahashi}},\
  and\ \bibinfo {author} {\bibfnamefont {S.}~\bibnamefont {Fujimoto}},\ }\href
  {https://doi.org/10.1103/PhysRevLett.103.020401} {\bibfield  {journal}
  {\bibinfo  {journal} {Phys. Rev. Lett.}\ }\textbf {\bibinfo {volume} {103}},\
  \bibinfo {pages} {020401} (\bibinfo {year} {2009})}\BibitemShut {NoStop}%
\bibitem [{\citenamefont {Qi}\ \emph {et~al.}(2010{\natexlab{a}})\citenamefont
  {Qi}, \citenamefont {Hughes},\ and\ \citenamefont {Zhang}}]{qi2010chiral}%
  \BibitemOpen
  \bibfield  {author} {\bibinfo {author} {\bibfnamefont {X.-L.}\ \bibnamefont
  {Qi}}, \bibinfo {author} {\bibfnamefont {T.~L.}\ \bibnamefont {Hughes}},\
  and\ \bibinfo {author} {\bibfnamefont {S.-C.}\ \bibnamefont {Zhang}},\ }\href
  {https://doi.org/10.1103/PhysRevB.82.184516} {\bibfield  {journal} {\bibinfo
  {journal} {Phys. Rev. B}\ }\textbf {\bibinfo {volume} {82}},\ \bibinfo
  {pages} {184516} (\bibinfo {year} {2010}{\natexlab{a}})}\BibitemShut
  {NoStop}%
\bibitem [{\citenamefont {Alicea}(2012)}]{alicea2012new}%
  \BibitemOpen
  \bibfield  {author} {\bibinfo {author} {\bibfnamefont {J.}~\bibnamefont
  {Alicea}},\ }\href {https://doi.org/10.1088/0034-4885/75/7/076501} {\bibfield
   {journal} {\bibinfo  {journal} {Reports on progress in physics}\ }\textbf
  {\bibinfo {volume} {75}},\ \bibinfo {pages} {076501} (\bibinfo {year}
  {2012})}\BibitemShut {NoStop}%
\bibitem [{\citenamefont {Niu}\ \emph {et~al.}(1985)\citenamefont {Niu},
  \citenamefont {Thouless},\ and\ \citenamefont {Wu}}]{niu1985quantized}%
  \BibitemOpen
  \bibfield  {author} {\bibinfo {author} {\bibfnamefont {Q.}~\bibnamefont
  {Niu}}, \bibinfo {author} {\bibfnamefont {D.~J.}\ \bibnamefont {Thouless}},\
  and\ \bibinfo {author} {\bibfnamefont {Y.-S.}\ \bibnamefont {Wu}},\ }\href
  {https://doi.org/10.1103/PhysRevB.31.3372} {\bibfield  {journal} {\bibinfo
  {journal} {Phys. Rev. B}\ }\textbf {\bibinfo {volume} {31}},\ \bibinfo
  {pages} {3372} (\bibinfo {year} {1985})}\BibitemShut {NoStop}%
\bibitem [{\citenamefont {Laughlin}(1981)}]{laughlin1981pumping}%
  \BibitemOpen
  \bibfield  {author} {\bibinfo {author} {\bibfnamefont {R.~B.}\ \bibnamefont
  {Laughlin}},\ }\href {https://doi.org/10.1103/PhysRevB.23.5632} {\bibfield
  {journal} {\bibinfo  {journal} {Phys. Rev. B}\ }\textbf {\bibinfo {volume}
  {23}},\ \bibinfo {pages} {5632} (\bibinfo {year} {1981})}\BibitemShut
  {NoStop}%
\bibitem [{\citenamefont {Thouless}(1983)}]{thouless1983quantization}%
  \BibitemOpen
  \bibfield  {author} {\bibinfo {author} {\bibfnamefont {D.~J.}\ \bibnamefont
  {Thouless}},\ }\href {https://doi.org/10.1103/PhysRevB.27.6083} {\bibfield
  {journal} {\bibinfo  {journal} {Phys. Rev. B}\ }\textbf {\bibinfo {volume}
  {27}},\ \bibinfo {pages} {6083} (\bibinfo {year} {1983})}\BibitemShut
  {NoStop}%
\bibitem [{\citenamefont {King-Smith}\ and\ \citenamefont
  {Vanderbilt}(1993)}]{vanderbilt1993polarization}%
  \BibitemOpen
  \bibfield  {author} {\bibinfo {author} {\bibfnamefont {R.~D.}\ \bibnamefont
  {King-Smith}}\ and\ \bibinfo {author} {\bibfnamefont {D.}~\bibnamefont
  {Vanderbilt}},\ }\href {https://doi.org/10.1103/PhysRevB.47.1651} {\bibfield
  {journal} {\bibinfo  {journal} {Phys. Rev. B}\ }\textbf {\bibinfo {volume}
  {47}},\ \bibinfo {pages} {1651} (\bibinfo {year} {1993})}\BibitemShut
  {NoStop}%
\bibitem [{\citenamefont {Thonhauser}\ and\ \citenamefont
  {Vanderbilt}(2006)}]{vanderbilt2006wannier}%
  \BibitemOpen
  \bibfield  {author} {\bibinfo {author} {\bibfnamefont {T.}~\bibnamefont
  {Thonhauser}}\ and\ \bibinfo {author} {\bibfnamefont {D.}~\bibnamefont
  {Vanderbilt}},\ }\href {https://doi.org/10.1103/PhysRevB.74.235111}
  {\bibfield  {journal} {\bibinfo  {journal} {Phys. Rev. B}\ }\textbf {\bibinfo
  {volume} {74}},\ \bibinfo {pages} {235111} (\bibinfo {year}
  {2006})}\BibitemShut {NoStop}%
\bibitem [{\citenamefont {Fu}\ and\ \citenamefont {Kane}(2006)}]{fu2006time}%
  \BibitemOpen
  \bibfield  {author} {\bibinfo {author} {\bibfnamefont {L.}~\bibnamefont
  {Fu}}\ and\ \bibinfo {author} {\bibfnamefont {C.~L.}\ \bibnamefont {Kane}},\
  }\href {https://doi.org/10.1103/PhysRevB.74.195312} {\bibfield  {journal}
  {\bibinfo  {journal} {Phys. Rev. B}\ }\textbf {\bibinfo {volume} {74}},\
  \bibinfo {pages} {195312} (\bibinfo {year} {2006})}\BibitemShut {NoStop}%
\bibitem [{\citenamefont {Qi}\ \emph {et~al.}(2009)\citenamefont {Qi},
  \citenamefont {Hughes}, \citenamefont {Raghu},\ and\ \citenamefont
  {Zhang}}]{qi2009time}%
  \BibitemOpen
  \bibfield  {author} {\bibinfo {author} {\bibfnamefont {X.-L.}\ \bibnamefont
  {Qi}}, \bibinfo {author} {\bibfnamefont {T.~L.}\ \bibnamefont {Hughes}},
  \bibinfo {author} {\bibfnamefont {S.}~\bibnamefont {Raghu}},\ and\ \bibinfo
  {author} {\bibfnamefont {S.-C.}\ \bibnamefont {Zhang}},\ }\href
  {https://doi.org/10.1103/PhysRevLett.102.187001} {\bibfield  {journal}
  {\bibinfo  {journal} {Phys. Rev. Lett.}\ }\textbf {\bibinfo {volume} {102}},\
  \bibinfo {pages} {187001} (\bibinfo {year} {2009})}\BibitemShut {NoStop}%
\bibitem [{\citenamefont {Qi}\ \emph {et~al.}(2010{\natexlab{b}})\citenamefont
  {Qi}, \citenamefont {Hughes},\ and\ \citenamefont
  {Zhang}}]{qi2010topological}%
  \BibitemOpen
  \bibfield  {author} {\bibinfo {author} {\bibfnamefont {X.-L.}\ \bibnamefont
  {Qi}}, \bibinfo {author} {\bibfnamefont {T.~L.}\ \bibnamefont {Hughes}},\
  and\ \bibinfo {author} {\bibfnamefont {S.-C.}\ \bibnamefont {Zhang}},\ }\href
  {https://doi.org/10.1103/PhysRevB.81.134508} {\bibfield  {journal} {\bibinfo
  {journal} {Phys. Rev. B}\ }\textbf {\bibinfo {volume} {81}},\ \bibinfo
  {pages} {134508} (\bibinfo {year} {2010}{\natexlab{b}})}\BibitemShut
  {NoStop}%
\bibitem [{\citenamefont {Zhang}\ \emph
  {et~al.}(2013{\natexlab{a}})\citenamefont {Zhang}, \citenamefont {Kane},\
  and\ \citenamefont {Mele}}]{zhang2013time}%
  \BibitemOpen
  \bibfield  {author} {\bibinfo {author} {\bibfnamefont {F.}~\bibnamefont
  {Zhang}}, \bibinfo {author} {\bibfnamefont {C.~L.}\ \bibnamefont {Kane}},\
  and\ \bibinfo {author} {\bibfnamefont {E.~J.}\ \bibnamefont {Mele}},\ }\href
  {https://doi.org/10.1103/PhysRevLett.111.056402} {\bibfield  {journal}
  {\bibinfo  {journal} {Phys. Rev. Lett.}\ }\textbf {\bibinfo {volume} {111}},\
  \bibinfo {pages} {056402} (\bibinfo {year} {2013}{\natexlab{a}})}\BibitemShut
  {NoStop}%
\bibitem [{\citenamefont {Zhang}\ and\ \citenamefont
  {Das~Sarma}(2021)}]{zhang2021intrinsic}%
  \BibitemOpen
  \bibfield  {author} {\bibinfo {author} {\bibfnamefont {R.-X.}\ \bibnamefont
  {Zhang}}\ and\ \bibinfo {author} {\bibfnamefont {S.}~\bibnamefont
  {Das~Sarma}},\ }\href {https://doi.org/10.1103/PhysRevLett.126.137001}
  {\bibfield  {journal} {\bibinfo  {journal} {Phys. Rev. Lett.}\ }\textbf
  {\bibinfo {volume} {126}},\ \bibinfo {pages} {137001} (\bibinfo {year}
  {2021})}\BibitemShut {NoStop}%
\bibitem [{\citenamefont {Zhang}\ \emph
  {et~al.}(2013{\natexlab{b}})\citenamefont {Zhang}, \citenamefont {Kane},\
  and\ \citenamefont {Mele}}]{zhang2013topo}%
  \BibitemOpen
  \bibfield  {author} {\bibinfo {author} {\bibfnamefont {F.}~\bibnamefont
  {Zhang}}, \bibinfo {author} {\bibfnamefont {C.~L.}\ \bibnamefont {Kane}},\
  and\ \bibinfo {author} {\bibfnamefont {E.~J.}\ \bibnamefont {Mele}},\ }\href
  {https://doi.org/10.1103/PhysRevLett.111.056403} {\bibfield  {journal}
  {\bibinfo  {journal} {Phys. Rev. Lett.}\ }\textbf {\bibinfo {volume} {111}},\
  \bibinfo {pages} {056403} (\bibinfo {year} {2013}{\natexlab{b}})}\BibitemShut
  {NoStop}%
\bibitem [{\citenamefont {Langbehn}\ \emph {et~al.}(2017)\citenamefont
  {Langbehn}, \citenamefont {Peng}, \citenamefont {Trifunovic}, \citenamefont
  {von Oppen},\ and\ \citenamefont {Brouwer}}]{langbehn2017reflection}%
  \BibitemOpen
  \bibfield  {author} {\bibinfo {author} {\bibfnamefont {J.}~\bibnamefont
  {Langbehn}}, \bibinfo {author} {\bibfnamefont {Y.}~\bibnamefont {Peng}},
  \bibinfo {author} {\bibfnamefont {L.}~\bibnamefont {Trifunovic}}, \bibinfo
  {author} {\bibfnamefont {F.}~\bibnamefont {von Oppen}},\ and\ \bibinfo
  {author} {\bibfnamefont {P.~W.}\ \bibnamefont {Brouwer}},\ }\href
  {https://doi.org/10.1103/PhysRevLett.119.246401} {\bibfield  {journal}
  {\bibinfo  {journal} {Phys. Rev. Lett.}\ }\textbf {\bibinfo {volume} {119}},\
  \bibinfo {pages} {246401} (\bibinfo {year} {2017})}\BibitemShut {NoStop}%
\bibitem [{\citenamefont {Meng}\ and\ \citenamefont
  {Balents}(2012)}]{meng2012weyl}%
  \BibitemOpen
  \bibfield  {author} {\bibinfo {author} {\bibfnamefont {T.}~\bibnamefont
  {Meng}}\ and\ \bibinfo {author} {\bibfnamefont {L.}~\bibnamefont {Balents}},\
  }\href {https://doi.org/10.1103/PhysRevB.86.054504} {\bibfield  {journal}
  {\bibinfo  {journal} {Phys. Rev. B}\ }\textbf {\bibinfo {volume} {86}},\
  \bibinfo {pages} {054504} (\bibinfo {year} {2012})}\BibitemShut {NoStop}%
\bibitem [{\citenamefont {Yang}\ \emph {et~al.}(2014)\citenamefont {Yang},
  \citenamefont {Pan},\ and\ \citenamefont {Zhang}}]{yang2014dirac}%
  \BibitemOpen
  \bibfield  {author} {\bibinfo {author} {\bibfnamefont {S.~A.}\ \bibnamefont
  {Yang}}, \bibinfo {author} {\bibfnamefont {H.}~\bibnamefont {Pan}},\ and\
  \bibinfo {author} {\bibfnamefont {F.}~\bibnamefont {Zhang}},\ }\href
  {https://doi.org/10.1103/PhysRevLett.113.046401} {\bibfield  {journal}
  {\bibinfo  {journal} {Phys. Rev. Lett.}\ }\textbf {\bibinfo {volume} {113}},\
  \bibinfo {pages} {046401} (\bibinfo {year} {2014})}\BibitemShut {NoStop}%
\bibitem [{\citenamefont {Kobayashi}\ and\ \citenamefont
  {Sato}(2015)}]{kobayashi2015dirac}%
  \BibitemOpen
  \bibfield  {author} {\bibinfo {author} {\bibfnamefont {S.}~\bibnamefont
  {Kobayashi}}\ and\ \bibinfo {author} {\bibfnamefont {M.}~\bibnamefont
  {Sato}},\ }\href {https://doi.org/10.1103/PhysRevLett.115.187001} {\bibfield
  {journal} {\bibinfo  {journal} {Phys. Rev. Lett.}\ }\textbf {\bibinfo
  {volume} {115}},\ \bibinfo {pages} {187001} (\bibinfo {year}
  {2015})}\BibitemShut {NoStop}%
\bibitem [{\citenamefont {Teo}\ and\ \citenamefont
  {Hughes}(2013)}]{teo2013existence}%
  \BibitemOpen
  \bibfield  {author} {\bibinfo {author} {\bibfnamefont {J.~C.~Y.}\
  \bibnamefont {Teo}}\ and\ \bibinfo {author} {\bibfnamefont {T.~L.}\
  \bibnamefont {Hughes}},\ }\href
  {https://doi.org/10.1103/PhysRevLett.111.047006} {\bibfield  {journal}
  {\bibinfo  {journal} {Phys. Rev. Lett.}\ }\textbf {\bibinfo {volume} {111}},\
  \bibinfo {pages} {047006} (\bibinfo {year} {2013})}\BibitemShut {NoStop}%
\bibitem [{\citenamefont {Benalcazar}\ \emph {et~al.}(2014)\citenamefont
  {Benalcazar}, \citenamefont {Teo},\ and\ \citenamefont
  {Hughes}}]{benalcazar2014class}%
  \BibitemOpen
  \bibfield  {author} {\bibinfo {author} {\bibfnamefont {W.~A.}\ \bibnamefont
  {Benalcazar}}, \bibinfo {author} {\bibfnamefont {J.~C.~Y.}\ \bibnamefont
  {Teo}},\ and\ \bibinfo {author} {\bibfnamefont {T.~L.}\ \bibnamefont
  {Hughes}},\ }\href {https://doi.org/10.1103/PhysRevB.89.224503} {\bibfield
  {journal} {\bibinfo  {journal} {Phys. Rev. B}\ }\textbf {\bibinfo {volume}
  {89}},\ \bibinfo {pages} {224503} (\bibinfo {year} {2014})}\BibitemShut
  {NoStop}%
\bibitem [{\citenamefont {Wang}\ \emph
  {et~al.}(2018{\natexlab{a}})\citenamefont {Wang}, \citenamefont {Lin},\ and\
  \citenamefont {Hughes}}]{wang2018weak}%
  \BibitemOpen
  \bibfield  {author} {\bibinfo {author} {\bibfnamefont {Y.}~\bibnamefont
  {Wang}}, \bibinfo {author} {\bibfnamefont {M.}~\bibnamefont {Lin}},\ and\
  \bibinfo {author} {\bibfnamefont {T.~L.}\ \bibnamefont {Hughes}},\ }\href
  {https://doi.org/10.1103/PhysRevB.98.165144} {\bibfield  {journal} {\bibinfo
  {journal} {Phys. Rev. B}\ }\textbf {\bibinfo {volume} {98}},\ \bibinfo
  {pages} {165144} (\bibinfo {year} {2018}{\natexlab{a}})}\BibitemShut
  {NoStop}%
\bibitem [{\citenamefont {Shapourian}\ \emph {et~al.}(2018)\citenamefont
  {Shapourian}, \citenamefont {Wang},\ and\ \citenamefont
  {Ryu}}]{shapourian2018topo}%
  \BibitemOpen
  \bibfield  {author} {\bibinfo {author} {\bibfnamefont {H.}~\bibnamefont
  {Shapourian}}, \bibinfo {author} {\bibfnamefont {Y.}~\bibnamefont {Wang}},\
  and\ \bibinfo {author} {\bibfnamefont {S.}~\bibnamefont {Ryu}},\ }\href
  {https://doi.org/10.1103/PhysRevB.97.094508} {\bibfield  {journal} {\bibinfo
  {journal} {Phys. Rev. B}\ }\textbf {\bibinfo {volume} {97}},\ \bibinfo
  {pages} {094508} (\bibinfo {year} {2018})}\BibitemShut {NoStop}%
\bibitem [{\citenamefont {Wang}\ \emph
  {et~al.}(2018{\natexlab{b}})\citenamefont {Wang}, \citenamefont {Liu},
  \citenamefont {Lu},\ and\ \citenamefont {Zhang}}]{wang2018high}%
  \BibitemOpen
  \bibfield  {author} {\bibinfo {author} {\bibfnamefont {Q.}~\bibnamefont
  {Wang}}, \bibinfo {author} {\bibfnamefont {C.-C.}\ \bibnamefont {Liu}},
  \bibinfo {author} {\bibfnamefont {Y.-M.}\ \bibnamefont {Lu}},\ and\ \bibinfo
  {author} {\bibfnamefont {F.}~\bibnamefont {Zhang}},\ }\href
  {https://doi.org/10.1103/PhysRevLett.121.186801} {\bibfield  {journal}
  {\bibinfo  {journal} {Phys. Rev. Lett.}\ }\textbf {\bibinfo {volume} {121}},\
  \bibinfo {pages} {186801} (\bibinfo {year} {2018}{\natexlab{b}})}\BibitemShut
  {NoStop}%
\bibitem [{\citenamefont {Yan}\ \emph {et~al.}(2018)\citenamefont {Yan},
  \citenamefont {Song},\ and\ \citenamefont {Wang}}]{yan2018majorana}%
  \BibitemOpen
  \bibfield  {author} {\bibinfo {author} {\bibfnamefont {Z.}~\bibnamefont
  {Yan}}, \bibinfo {author} {\bibfnamefont {F.}~\bibnamefont {Song}},\ and\
  \bibinfo {author} {\bibfnamefont {Z.}~\bibnamefont {Wang}},\ }\href
  {https://doi.org/10.1103/PhysRevLett.121.096803} {\bibfield  {journal}
  {\bibinfo  {journal} {Phys. Rev. Lett.}\ }\textbf {\bibinfo {volume} {121}},\
  \bibinfo {pages} {096803} (\bibinfo {year} {2018})}\BibitemShut {NoStop}%
\bibitem [{\citenamefont {You}\ \emph {et~al.}(2018)\citenamefont {You},
  \citenamefont {Devakul}, \citenamefont {Burnell},\ and\ \citenamefont
  {Neupert}}]{you2018higher}%
  \BibitemOpen
  \bibfield  {author} {\bibinfo {author} {\bibfnamefont {Y.}~\bibnamefont
  {You}}, \bibinfo {author} {\bibfnamefont {T.}~\bibnamefont {Devakul}},
  \bibinfo {author} {\bibfnamefont {F.~J.}\ \bibnamefont {Burnell}},\ and\
  \bibinfo {author} {\bibfnamefont {T.}~\bibnamefont {Neupert}},\ }\href
  {https://doi.org/10.1103/PhysRevB.98.235102} {\bibfield  {journal} {\bibinfo
  {journal} {Phys. Rev. B}\ }\textbf {\bibinfo {volume} {98}},\ \bibinfo
  {pages} {235102} (\bibinfo {year} {2018})}\BibitemShut {NoStop}%
\bibitem [{\citenamefont {Zhang}\ \emph
  {et~al.}(2019{\natexlab{a}})\citenamefont {Zhang}, \citenamefont {Cole},\
  and\ \citenamefont {Das~Sarma}}]{zhang2019helical}%
  \BibitemOpen
  \bibfield  {author} {\bibinfo {author} {\bibfnamefont {R.-X.}\ \bibnamefont
  {Zhang}}, \bibinfo {author} {\bibfnamefont {W.~S.}\ \bibnamefont {Cole}},\
  and\ \bibinfo {author} {\bibfnamefont {S.}~\bibnamefont {Das~Sarma}},\ }\href
  {https://doi.org/10.1103/PhysRevLett.122.187001} {\bibfield  {journal}
  {\bibinfo  {journal} {Phys. Rev. Lett.}\ }\textbf {\bibinfo {volume} {122}},\
  \bibinfo {pages} {187001} (\bibinfo {year} {2019}{\natexlab{a}})}\BibitemShut
  {NoStop}%
\bibitem [{\citenamefont {Ghorashi}\ \emph {et~al.}(2019)\citenamefont
  {Ghorashi}, \citenamefont {Hu}, \citenamefont {Hughes},\ and\ \citenamefont
  {Rossi}}]{ghorashi2019second}%
  \BibitemOpen
  \bibfield  {author} {\bibinfo {author} {\bibfnamefont {S.~A.~A.}\
  \bibnamefont {Ghorashi}}, \bibinfo {author} {\bibfnamefont {X.}~\bibnamefont
  {Hu}}, \bibinfo {author} {\bibfnamefont {T.~L.}\ \bibnamefont {Hughes}},\
  and\ \bibinfo {author} {\bibfnamefont {E.}~\bibnamefont {Rossi}},\ }\href
  {https://doi.org/10.1103/PhysRevB.100.020509} {\bibfield  {journal} {\bibinfo
   {journal} {Phys. Rev. B}\ }\textbf {\bibinfo {volume} {100}},\ \bibinfo
  {pages} {020509} (\bibinfo {year} {2019})}\BibitemShut {NoStop}%
\bibitem [{\citenamefont {Zhang}\ \emph
  {et~al.}(2019{\natexlab{b}})\citenamefont {Zhang}, \citenamefont {Cole},
  \citenamefont {Wu},\ and\ \citenamefont {Das~Sarma}}]{zhang2019higher}%
  \BibitemOpen
  \bibfield  {author} {\bibinfo {author} {\bibfnamefont {R.-X.}\ \bibnamefont
  {Zhang}}, \bibinfo {author} {\bibfnamefont {W.~S.}\ \bibnamefont {Cole}},
  \bibinfo {author} {\bibfnamefont {X.}~\bibnamefont {Wu}},\ and\ \bibinfo
  {author} {\bibfnamefont {S.}~\bibnamefont {Das~Sarma}},\ }\href
  {https://doi.org/10.1103/PhysRevLett.123.167001} {\bibfield  {journal}
  {\bibinfo  {journal} {Phys. Rev. Lett.}\ }\textbf {\bibinfo {volume} {123}},\
  \bibinfo {pages} {167001} (\bibinfo {year} {2019}{\natexlab{b}})}\BibitemShut
  {NoStop}%
\bibitem [{\citenamefont {Volpez}\ \emph {et~al.}(2019)\citenamefont {Volpez},
  \citenamefont {Loss},\ and\ \citenamefont {Klinovaja}}]{volpez2019second}%
  \BibitemOpen
  \bibfield  {author} {\bibinfo {author} {\bibfnamefont {Y.}~\bibnamefont
  {Volpez}}, \bibinfo {author} {\bibfnamefont {D.}~\bibnamefont {Loss}},\ and\
  \bibinfo {author} {\bibfnamefont {J.}~\bibnamefont {Klinovaja}},\ }\href
  {https://doi.org/10.1103/PhysRevLett.122.126402} {\bibfield  {journal}
  {\bibinfo  {journal} {Phys. Rev. Lett.}\ }\textbf {\bibinfo {volume} {122}},\
  \bibinfo {pages} {126402} (\bibinfo {year} {2019})}\BibitemShut {NoStop}%
\bibitem [{\citenamefont {Liu}\ \emph {et~al.}(2018)\citenamefont {Liu},
  \citenamefont {He},\ and\ \citenamefont {Nori}}]{liu2018majorana}%
  \BibitemOpen
  \bibfield  {author} {\bibinfo {author} {\bibfnamefont {T.}~\bibnamefont
  {Liu}}, \bibinfo {author} {\bibfnamefont {J.~J.}\ \bibnamefont {He}},\ and\
  \bibinfo {author} {\bibfnamefont {F.}~\bibnamefont {Nori}},\ }\href
  {https://doi.org/10.1103/PhysRevB.98.245413} {\bibfield  {journal} {\bibinfo
  {journal} {Phys. Rev. B}\ }\textbf {\bibinfo {volume} {98}},\ \bibinfo
  {pages} {245413} (\bibinfo {year} {2018})}\BibitemShut {NoStop}%
\bibitem [{\citenamefont {Yan}(2019)}]{yan2019higher}%
  \BibitemOpen
  \bibfield  {author} {\bibinfo {author} {\bibfnamefont {Z.}~\bibnamefont
  {Yan}},\ }\href {https://doi.org/10.1103/PhysRevLett.123.177001} {\bibfield
  {journal} {\bibinfo  {journal} {Phys. Rev. Lett.}\ }\textbf {\bibinfo
  {volume} {123}},\ \bibinfo {pages} {177001} (\bibinfo {year}
  {2019})}\BibitemShut {NoStop}%
\bibitem [{\citenamefont {Hsu}\ \emph {et~al.}(2020)\citenamefont {Hsu},
  \citenamefont {Cole}, \citenamefont {Zhang},\ and\ \citenamefont
  {Sau}}]{hsu2020inversion}%
  \BibitemOpen
  \bibfield  {author} {\bibinfo {author} {\bibfnamefont {Y.-T.}\ \bibnamefont
  {Hsu}}, \bibinfo {author} {\bibfnamefont {W.~S.}\ \bibnamefont {Cole}},
  \bibinfo {author} {\bibfnamefont {R.-X.}\ \bibnamefont {Zhang}},\ and\
  \bibinfo {author} {\bibfnamefont {J.~D.}\ \bibnamefont {Sau}},\ }\href
  {https://doi.org/10.1103/PhysRevLett.125.097001} {\bibfield  {journal}
  {\bibinfo  {journal} {Phys. Rev. Lett.}\ }\textbf {\bibinfo {volume} {125}},\
  \bibinfo {pages} {097001} (\bibinfo {year} {2020})}\BibitemShut {NoStop}%
\bibitem [{\citenamefont {Pan}\ \emph {et~al.}(2019)\citenamefont {Pan},
  \citenamefont {Yang}, \citenamefont {Chen}, \citenamefont {Xu}, \citenamefont
  {Liu},\ and\ \citenamefont {Liu}}]{pan2019lattice}%
  \BibitemOpen
  \bibfield  {author} {\bibinfo {author} {\bibfnamefont {X.-H.}\ \bibnamefont
  {Pan}}, \bibinfo {author} {\bibfnamefont {K.-J.}\ \bibnamefont {Yang}},
  \bibinfo {author} {\bibfnamefont {L.}~\bibnamefont {Chen}}, \bibinfo {author}
  {\bibfnamefont {G.}~\bibnamefont {Xu}}, \bibinfo {author} {\bibfnamefont
  {C.-X.}\ \bibnamefont {Liu}},\ and\ \bibinfo {author} {\bibfnamefont
  {X.}~\bibnamefont {Liu}},\ }\href
  {https://doi.org/10.1103/PhysRevLett.123.156801} {\bibfield  {journal}
  {\bibinfo  {journal} {Phys. Rev. Lett.}\ }\textbf {\bibinfo {volume} {123}},\
  \bibinfo {pages} {156801} (\bibinfo {year} {2019})}\BibitemShut {NoStop}%
\bibitem [{\citenamefont {Zhang}\ \emph
  {et~al.}(2020{\natexlab{a}})\citenamefont {Zhang}, \citenamefont {Sau},\ and\
  \citenamefont {Sarma}}]{zhang2020kitaev}%
  \BibitemOpen
  \bibfield  {author} {\bibinfo {author} {\bibfnamefont {R.-X.}\ \bibnamefont
  {Zhang}}, \bibinfo {author} {\bibfnamefont {J.~D.}\ \bibnamefont {Sau}},\
  and\ \bibinfo {author} {\bibfnamefont {S.~D.}\ \bibnamefont {Sarma}},\ }\href
  {https://arxiv.org/abs/2003.02559} {\bibfield  {journal} {\bibinfo  {journal}
  {arXiv:2003.02559}\ } (\bibinfo {year} {2020}{\natexlab{a}})}\BibitemShut
  {NoStop}%
\bibitem [{\citenamefont {Zhang}(2022)}]{zhang2022strongly}%
  \BibitemOpen
  \bibfield  {author} {\bibinfo {author} {\bibfnamefont {J.-H.}\ \bibnamefont
  {Zhang}},\ }\href {https://doi.org/10.1103/PhysRevB.106.L020503} {\bibfield
  {journal} {\bibinfo  {journal} {Phys. Rev. B}\ }\textbf {\bibinfo {volume}
  {106}},\ \bibinfo {pages} {L020503} (\bibinfo {year} {2022})}\BibitemShut
  {NoStop}%
\bibitem [{\citenamefont {Hu}\ and\ \citenamefont
  {Zhang}(2022{\natexlab{a}})}]{hu2022higher}%
  \BibitemOpen
  \bibfield  {author} {\bibinfo {author} {\bibfnamefont {L.-H.}\ \bibnamefont
  {Hu}}\ and\ \bibinfo {author} {\bibfnamefont {R.-X.}\ \bibnamefont {Zhang}},\
  }\href {https://arxiv.org/abs/2207.10113} {\bibfield  {journal} {\bibinfo
  {journal} {arXiv:2207.10113}\ } (\bibinfo {year}
  {2022}{\natexlab{a}})}\BibitemShut {NoStop}%
\bibitem [{\citenamefont {Khalaf}(2018)}]{khalaf2018higher}%
  \BibitemOpen
  \bibfield  {author} {\bibinfo {author} {\bibfnamefont {E.}~\bibnamefont
  {Khalaf}},\ }\href {https://doi.org/10.1103/PhysRevB.97.205136} {\bibfield
  {journal} {\bibinfo  {journal} {Phys. Rev. B}\ }\textbf {\bibinfo {volume}
  {97}},\ \bibinfo {pages} {205136} (\bibinfo {year} {2018})}\BibitemShut
  {NoStop}%
\bibitem [{\citenamefont {Trifunovic}\ and\ \citenamefont
  {Brouwer}(2019)}]{trifunovic2019higher}%
  \BibitemOpen
  \bibfield  {author} {\bibinfo {author} {\bibfnamefont {L.}~\bibnamefont
  {Trifunovic}}\ and\ \bibinfo {author} {\bibfnamefont {P.~W.}\ \bibnamefont
  {Brouwer}},\ }\href {https://doi.org/10.1103/PhysRevX.9.011012} {\bibfield
  {journal} {\bibinfo  {journal} {Phys. Rev. X}\ }\textbf {\bibinfo {volume}
  {9}},\ \bibinfo {pages} {011012} (\bibinfo {year} {2019})}\BibitemShut
  {NoStop}%
\bibitem [{\citenamefont {Skurativska}\ \emph {et~al.}(2020)\citenamefont
  {Skurativska}, \citenamefont {Neupert},\ and\ \citenamefont
  {Fischer}}]{titus2020atomic}%
  \BibitemOpen
  \bibfield  {author} {\bibinfo {author} {\bibfnamefont {A.}~\bibnamefont
  {Skurativska}}, \bibinfo {author} {\bibfnamefont {T.}~\bibnamefont
  {Neupert}},\ and\ \bibinfo {author} {\bibfnamefont {M.~H.}\ \bibnamefont
  {Fischer}},\ }\href {https://doi.org/10.1103/PhysRevResearch.2.013064}
  {\bibfield  {journal} {\bibinfo  {journal} {Phys. Rev. Research}\ }\textbf
  {\bibinfo {volume} {2}},\ \bibinfo {pages} {013064} (\bibinfo {year}
  {2020})}\BibitemShut {NoStop}%
\bibitem [{\citenamefont {Roberts}\ \emph {et~al.}(2020)\citenamefont
  {Roberts}, \citenamefont {Behrends},\ and\ \citenamefont
  {B\'eri}}]{roberts2020second}%
  \BibitemOpen
  \bibfield  {author} {\bibinfo {author} {\bibfnamefont {E.}~\bibnamefont
  {Roberts}}, \bibinfo {author} {\bibfnamefont {J.}~\bibnamefont {Behrends}},\
  and\ \bibinfo {author} {\bibfnamefont {B.}~\bibnamefont {B\'eri}},\ }\href
  {https://doi.org/10.1103/PhysRevB.101.155133} {\bibfield  {journal} {\bibinfo
   {journal} {Phys. Rev. B}\ }\textbf {\bibinfo {volume} {101}},\ \bibinfo
  {pages} {155133} (\bibinfo {year} {2020})}\BibitemShut {NoStop}%
\bibitem [{\citenamefont {Wu}\ and\ \citenamefont
  {Martin}(2017)}]{wu2017majorana}%
  \BibitemOpen
  \bibfield  {author} {\bibinfo {author} {\bibfnamefont {F.}~\bibnamefont
  {Wu}}\ and\ \bibinfo {author} {\bibfnamefont {I.}~\bibnamefont {Martin}},\
  }\href {https://doi.org/10.1103/PhysRevB.95.224503} {\bibfield  {journal}
  {\bibinfo  {journal} {Phys. Rev. B}\ }\textbf {\bibinfo {volume} {95}},\
  \bibinfo {pages} {224503} (\bibinfo {year} {2017})}\BibitemShut {NoStop}%
\bibitem [{\citenamefont {Fu}\ and\ \citenamefont {Kane}(2008)}]{fu2008vortex}%
  \BibitemOpen
  \bibfield  {author} {\bibinfo {author} {\bibfnamefont {L.}~\bibnamefont
  {Fu}}\ and\ \bibinfo {author} {\bibfnamefont {C.~L.}\ \bibnamefont {Kane}},\
  }\href {https://doi.org/10.1103/PhysRevLett.100.096407} {\bibfield  {journal}
  {\bibinfo  {journal} {Phys. Rev. Lett.}\ }\textbf {\bibinfo {volume} {100}},\
  \bibinfo {pages} {096407} (\bibinfo {year} {2008})}\BibitemShut {NoStop}%
\bibitem [{\citenamefont {Liu}\ \emph {et~al.}(2014)\citenamefont {Liu},
  \citenamefont {He},\ and\ \citenamefont {Law}}]{liu2014lattice}%
  \BibitemOpen
  \bibfield  {author} {\bibinfo {author} {\bibfnamefont {X.-J.}\ \bibnamefont
  {Liu}}, \bibinfo {author} {\bibfnamefont {J.~J.}\ \bibnamefont {He}},\ and\
  \bibinfo {author} {\bibfnamefont {K.~T.}\ \bibnamefont {Law}},\ }\href
  {https://doi.org/10.1103/PhysRevB.90.235141} {\bibfield  {journal} {\bibinfo
  {journal} {Phys. Rev. B}\ }\textbf {\bibinfo {volume} {90}},\ \bibinfo
  {pages} {235141} (\bibinfo {year} {2014})}\BibitemShut {NoStop}%
\bibitem [{\citenamefont {Hu}\ and\ \citenamefont
  {Zhang}(2022{\natexlab{b}})}]{hu2022topological}%
  \BibitemOpen
  \bibfield  {author} {\bibinfo {author} {\bibfnamefont {L.-H.}\ \bibnamefont
  {Hu}}\ and\ \bibinfo {author} {\bibfnamefont {R.-X.}\ \bibnamefont {Zhang}},\
  }\href {https://arxiv.org/abs/2204.03175} {\bibfield  {journal} {\bibinfo
  {journal} {arXiv:2204.03175}\ } (\bibinfo {year}
  {2022}{\natexlab{b}})}\BibitemShut {NoStop}%
\bibitem [{\citenamefont {Kopnin}\ and\ \citenamefont
  {Salomaa}(1991)}]{kopnin1991mutual}%
  \BibitemOpen
  \bibfield  {author} {\bibinfo {author} {\bibfnamefont {N.~B.}\ \bibnamefont
  {Kopnin}}\ and\ \bibinfo {author} {\bibfnamefont {M.~M.}\ \bibnamefont
  {Salomaa}},\ }\href {https://doi.org/10.1103/PhysRevB.44.9667} {\bibfield
  {journal} {\bibinfo  {journal} {Phys. Rev. B}\ }\textbf {\bibinfo {volume}
  {44}},\ \bibinfo {pages} {9667} (\bibinfo {year} {1991})}\BibitemShut
  {NoStop}%
\bibitem [{\citenamefont {Volovik}(1999)}]{volovik1999fermion}%
  \BibitemOpen
  \bibfield  {author} {\bibinfo {author} {\bibfnamefont {G.}~\bibnamefont
  {Volovik}},\ }\href {https://doi.org/10.1134/1.568223} {\bibfield  {journal}
  {\bibinfo  {journal} {Journal of Experimental and Theoretical Physics
  Letters}\ }\textbf {\bibinfo {volume} {70}},\ \bibinfo {pages} {609}
  (\bibinfo {year} {1999})}\BibitemShut {NoStop}%
\bibitem [{\citenamefont {Tewari}\ \emph
  {et~al.}(2007{\natexlab{a}})\citenamefont {Tewari}, \citenamefont
  {Das~Sarma},\ and\ \citenamefont {Lee}}]{sumanta2007index}%
  \BibitemOpen
  \bibfield  {author} {\bibinfo {author} {\bibfnamefont {S.}~\bibnamefont
  {Tewari}}, \bibinfo {author} {\bibfnamefont {S.}~\bibnamefont {Das~Sarma}},\
  and\ \bibinfo {author} {\bibfnamefont {D.-H.}\ \bibnamefont {Lee}},\ }\href
  {https://doi.org/10.1103/PhysRevLett.99.037001} {\bibfield  {journal}
  {\bibinfo  {journal} {Phys. Rev. Lett.}\ }\textbf {\bibinfo {volume} {99}},\
  \bibinfo {pages} {037001} (\bibinfo {year} {2007}{\natexlab{a}})}\BibitemShut
  {NoStop}%
\bibitem [{\citenamefont {Gurarie}\ and\ \citenamefont
  {Radzihovsky}(2007)}]{gurarie2007zero}%
  \BibitemOpen
  \bibfield  {author} {\bibinfo {author} {\bibfnamefont {V.}~\bibnamefont
  {Gurarie}}\ and\ \bibinfo {author} {\bibfnamefont {L.}~\bibnamefont
  {Radzihovsky}},\ }\href {https://doi.org/10.1103/PhysRevB.75.212509}
  {\bibfield  {journal} {\bibinfo  {journal} {Phys. Rev. B}\ }\textbf {\bibinfo
  {volume} {75}},\ \bibinfo {pages} {212509} (\bibinfo {year}
  {2007})}\BibitemShut {NoStop}%
\bibitem [{\citenamefont {Tewari}\ \emph
  {et~al.}(2007{\natexlab{b}})\citenamefont {Tewari}, \citenamefont
  {Das~Sarma}, \citenamefont {Nayak}, \citenamefont {Zhang},\ and\
  \citenamefont {Zoller}}]{sumanta2007quantum}%
  \BibitemOpen
  \bibfield  {author} {\bibinfo {author} {\bibfnamefont {S.}~\bibnamefont
  {Tewari}}, \bibinfo {author} {\bibfnamefont {S.}~\bibnamefont {Das~Sarma}},
  \bibinfo {author} {\bibfnamefont {C.}~\bibnamefont {Nayak}}, \bibinfo
  {author} {\bibfnamefont {C.}~\bibnamefont {Zhang}},\ and\ \bibinfo {author}
  {\bibfnamefont {P.}~\bibnamefont {Zoller}},\ }\href
  {https://doi.org/10.1103/PhysRevLett.98.010506} {\bibfield  {journal}
  {\bibinfo  {journal} {Phys. Rev. Lett.}\ }\textbf {\bibinfo {volume} {98}},\
  \bibinfo {pages} {010506} (\bibinfo {year} {2007}{\natexlab{b}})}\BibitemShut
  {NoStop}%
\bibitem [{\citenamefont {Cheng}\ \emph {et~al.}(2009)\citenamefont {Cheng},
  \citenamefont {Lutchyn}, \citenamefont {Galitski},\ and\ \citenamefont
  {Das~Sarma}}]{cheng2009splitting}%
  \BibitemOpen
  \bibfield  {author} {\bibinfo {author} {\bibfnamefont {M.}~\bibnamefont
  {Cheng}}, \bibinfo {author} {\bibfnamefont {R.~M.}\ \bibnamefont {Lutchyn}},
  \bibinfo {author} {\bibfnamefont {V.}~\bibnamefont {Galitski}},\ and\
  \bibinfo {author} {\bibfnamefont {S.}~\bibnamefont {Das~Sarma}},\ }\href
  {https://doi.org/10.1103/PhysRevLett.103.107001} {\bibfield  {journal}
  {\bibinfo  {journal} {Phys. Rev. Lett.}\ }\textbf {\bibinfo {volume} {103}},\
  \bibinfo {pages} {107001} (\bibinfo {year} {2009})}\BibitemShut {NoStop}%
\bibitem [{\citenamefont {Wieder}\ \emph {et~al.}(2020)\citenamefont {Wieder},
  \citenamefont {Wang}, \citenamefont {Cano}, \citenamefont {Dai},
  \citenamefont {Schoop}, \citenamefont {Bradlyn},\ and\ \citenamefont
  {Bernevig}}]{wieder2020strong}%
  \BibitemOpen
  \bibfield  {author} {\bibinfo {author} {\bibfnamefont {B.~J.}\ \bibnamefont
  {Wieder}}, \bibinfo {author} {\bibfnamefont {Z.}~\bibnamefont {Wang}},
  \bibinfo {author} {\bibfnamefont {J.}~\bibnamefont {Cano}}, \bibinfo {author}
  {\bibfnamefont {X.}~\bibnamefont {Dai}}, \bibinfo {author} {\bibfnamefont
  {L.~M.}\ \bibnamefont {Schoop}}, \bibinfo {author} {\bibfnamefont
  {B.}~\bibnamefont {Bradlyn}},\ and\ \bibinfo {author} {\bibfnamefont {B.~A.}\
  \bibnamefont {Bernevig}},\ }\href
  {https://doi.org/10.1038/s41467-020-14443-5} {\bibfield  {journal} {\bibinfo
  {journal} {Nature communications}\ }\textbf {\bibinfo {volume} {11}},\
  \bibinfo {pages} {1} (\bibinfo {year} {2020})}\BibitemShut {NoStop}%
\bibitem [{\citenamefont {Zhang}\ \emph
  {et~al.}(2020{\natexlab{b}})\citenamefont {Zhang}, \citenamefont {Hsu},\ and\
  \citenamefont {Das~Sarma}}]{zhang2020dirac}%
  \BibitemOpen
  \bibfield  {author} {\bibinfo {author} {\bibfnamefont {R.-X.}\ \bibnamefont
  {Zhang}}, \bibinfo {author} {\bibfnamefont {Y.-T.}\ \bibnamefont {Hsu}},\
  and\ \bibinfo {author} {\bibfnamefont {S.}~\bibnamefont {Das~Sarma}},\ }\href
  {https://doi.org/10.1103/PhysRevB.102.094503} {\bibfield  {journal} {\bibinfo
   {journal} {Phys. Rev. B}\ }\textbf {\bibinfo {volume} {102}},\ \bibinfo
  {pages} {094503} (\bibinfo {year} {2020}{\natexlab{b}})}\BibitemShut
  {NoStop}%
\bibitem [{\citenamefont {Fang}\ \emph {et~al.}(2014)\citenamefont {Fang},
  \citenamefont {Gilbert},\ and\ \citenamefont {Bernevig}}]{fang2014new}%
  \BibitemOpen
  \bibfield  {author} {\bibinfo {author} {\bibfnamefont {C.}~\bibnamefont
  {Fang}}, \bibinfo {author} {\bibfnamefont {M.~J.}\ \bibnamefont {Gilbert}},\
  and\ \bibinfo {author} {\bibfnamefont {B.~A.}\ \bibnamefont {Bernevig}},\
  }\href {https://doi.org/10.1103/PhysRevLett.112.106401} {\bibfield  {journal}
  {\bibinfo  {journal} {Phys. Rev. Lett.}\ }\textbf {\bibinfo {volume} {112}},\
  \bibinfo {pages} {106401} (\bibinfo {year} {2014})}\BibitemShut {NoStop}%
\bibitem [{\citenamefont {Zhang}\ and\ \citenamefont
  {Liu}(2015)}]{zhang2015magnetic}%
  \BibitemOpen
  \bibfield  {author} {\bibinfo {author} {\bibfnamefont {R.-X.}\ \bibnamefont
  {Zhang}}\ and\ \bibinfo {author} {\bibfnamefont {C.-X.}\ \bibnamefont
  {Liu}},\ }\href {https://doi.org/10.1103/PhysRevB.91.115317} {\bibfield
  {journal} {\bibinfo  {journal} {Phys. Rev. B}\ }\textbf {\bibinfo {volume}
  {91}},\ \bibinfo {pages} {115317} (\bibinfo {year} {2015})}\BibitemShut
  {NoStop}%
\bibitem [{\citenamefont {Li}\ and\ \citenamefont
  {Sun}(2020)}]{li2020pfaffian}%
  \BibitemOpen
  \bibfield  {author} {\bibinfo {author} {\bibfnamefont {H.}~\bibnamefont
  {Li}}\ and\ \bibinfo {author} {\bibfnamefont {K.}~\bibnamefont {Sun}},\
  }\href {https://doi.org/10.1103/PhysRevLett.124.036401} {\bibfield  {journal}
  {\bibinfo  {journal} {Phys. Rev. Lett.}\ }\textbf {\bibinfo {volume} {124}},\
  \bibinfo {pages} {036401} (\bibinfo {year} {2020})}\BibitemShut {NoStop}%
\bibitem [{\citenamefont {You}\ and\ \citenamefont
  {Hughes}()}]{you2022unpublished}%
  \BibitemOpen
  \bibfield  {author} {\bibinfo {author} {\bibfnamefont {Y.}~\bibnamefont
  {You}}\ and\ \bibinfo {author} {\bibfnamefont {T.~L.}\ \bibnamefont
  {Hughes}},\ }\href@noop {} {\bibinfo  {journal} {unpublished}\ }\BibitemShut
  {NoStop}%
\end{thebibliography}%

\appendix
\newpage
\onecolumngrid

\section{$C_n$ representation for MZMs}
It is important to understand how corner MZMs from each patch transform under $C_n$, with and without a $\mathbb{Z}_m$ vortex. Let us start with a collection of vortex-free patches with a constant, vanishing SC phase $\alpha=0$. Note that a corner MZM is generally described by $\gamma_{i}^{(S_j)} = \psi_{i}^{(S_j)}+(\psi_{i}^{(S_j)})^{\dagger}$, where $i,j\in\{1,2,...,n\}$ and the fermion operator is defined as $\psi_i^{(S_j)} = \int d^2 {\bf r}\sum_{l} u_l ({\bf r -r}_i^{(S_j)}) c_{{\bf r}, l}$. Here $c_{{\bf r},l}$ annihilates an electron at ${\bf r}$ that carries a spin/orbital index $l$  and ${\bf r}_i^{(S_j)}$ is the position for the $i$-th corner of $S_j$. Geometrically, it is clear that $C_n$ will relate MZMs of different patches in the following way:
\begin{equation}
	\gamma_{i}^{(S_j)} \xrightarrow{C_n} \gamma_{i+1}^{(S_{j+1})} \xrightarrow{C_n} ...  \xrightarrow{C_n} \gamma_{i+n-1}^{(S_{j+n-1})}  \xrightarrow{C_n} \gamma_{i}^{(S_{j})}
\end{equation}
up to some $U(1)$ phase factors, given that the small patches are identical to one another. Since $(C_n)^n=-1$ for spinful fermions, we have $C_n \psi_i^{(S_j)} C_n^{-1} = \omega_n \psi_{i+1}^{(S_{j+1})}$ with $\omega_n=e^{i\pi/n}$. We note that this gauge choice of $\omega_n$ is not unique, and an alternative option is discussed in Ref.~\cite{benalcazar2014class}. Therefore, we have 
\begin{equation}
	C_n \gamma_{i}^{(S_j)} C_n^{-1} = U_{\frac{2\pi}{n}} \gamma_{i+1}^{(S_{j+1})},\ \ \ \text{no vortex},
	\label{eq-Cn-MZM-no-vortex}
\end{equation}
with $U_{\varphi} = \text{diag}[e^{i\varphi/2} \mathbbm{1}_{N}, e^{-i\varphi/2}\mathbbm{1}_{N}]$. Here $\mathbbm{1}_N$ is an $N\times N$ identity matrix and $N$ is the matrix rank of $h_0({\bf k})$.

Introducing a quantum vortex will make MZMs effectively spinless under $C_n$. To see this, we first consider to update Eq.~\ref{eq-ham-H} to
\begin{equation}
	H^{(\alpha)}({\bf k}) = \begin{pmatrix}
		h_0({\bf k}) & \Delta({\bf k}) e^{i\alpha} \\
		\Delta({\bf k}) e^{-i\alpha} & -h(-{\bf k})_0^T \\
	\end{pmatrix},
	\label{eq-ham-Halpha}
\end{equation}
where the constant SC phase $\alpha\neq 0$. Eq.~\ref{eq-ham-Halpha} and Eq.~\ref{eq-ham-H} are related by the unitary matrix $U_{\varphi}$, with $H^{(\alpha)}({\bf k})= U_{\alpha} H({\bf k}) U_{\alpha}^{-1}$. Therefore, if $H({\bf k})$ hosts a MZM $\gamma_{{\bf r}_0}= \psi_{{\bf r}_0}  + \psi_{{\bf r}_0}^{\dagger} $ at ${\bf r}_0$, the same MZM will now be updated to
\begin{equation}
	\gamma^{(\alpha)}_{{\bf r}_0} =e^{i\alpha/2} \psi_{{\bf r}_0}  + e^{-i\alpha/2} \psi_{{\bf r}_0}^{\dagger} = U_{\alpha} \gamma_{{\bf r}_0},
\end{equation}
under the effect of $\alpha$. Consider a $\mathbb{Z}_n$ vortex formed by an $n$-patch $S_\Sigma$, with each $S_j$ locally hosting $H^{(\frac{2\pi}{n}j)}({\bf k})$. Note that the phase jump provided by the vortex configuration from $S_j$ to $S_{j+1}$ exactly coincides with the $C_n$ rotation phase in Eq.~\ref{eq-Cn-MZM-no-vortex}. Consequently, the rotating rule for MZMs will be updated to
\begin{equation}
	C_n \gamma_{i}^{(S_j)} C_n^{-1} = \gamma_{i+1}^{(S_{j+1})},\ \ \ \text{with vortex}.
\end{equation}
In general, we expect the MZMs to follow $C_n^n=-(-1)^v$, where $v$ is the vorticity of the SC vortex.

\section{Minimal Model for HOTSC$_n$}

In this section, we provide detailed discussions on the HOTSC$_n$ model $H_\text{BL}({\bf k})$.

\subsection{Expression of $h_c({\bf k})$}

As discussed in the main text, detailed forms of $h_c({\bf k})$ depend on the choice of basis function $\Psi_\text{BL}=(\Psi_{\nu}, \Psi_{\nu'})$. Clearly, $h_c({\bf k})$ is only relevant to $\nu-\nu'$. Namely, $(\nu,\nu')=(0,0)$ and $(\nu,\nu')=(\frac{n}{2},\frac{n}{2})$ share the same $h_c({\bf k})$. In this case, we only need to consider $(0,0)$ and $(0,\frac{n}{2})$. We denote $h_c({\bf k})$ for $(\nu,\nu')$ as $h_c^{(\nu,\nu')}({\bf k})$. Then it is easy to show that to satisfy $C_n$ symmetry, we have 
\begin{eqnarray}
   	h_c^{(0,\frac{n}{2})}({\bf k}) = \begin{pmatrix}
   	\alpha k_-^{\frac{n}{2}} & \beta k_+^{\frac{n}{2}-1}\\
   	\beta' k_-^{\frac{n}{2}-1} & \alpha' k_+^{\frac{n}{2}}  \\
   \end{pmatrix} + {\cal O}(k^{\frac{n}{2}+1}),\ \ \ 
	h_c^{(0,0)}({\bf k}) = \begin{pmatrix}
	A & B k_- \\
	B' k_+ & A' \\
\end{pmatrix} + {\cal O}(k^2).
\end{eqnarray}
Particle-hole symmetry further requires
\begin{eqnarray}
	A = - A',\ \ B=B',\ \ 
	\alpha = (-1)^{\frac{n}{2}+1} \alpha',\ \ \beta = (-1)^{\frac{n}{2}} \beta'.
\end{eqnarray}
We thus conclude that 

\begin{eqnarray}
	h_c^{(0,\frac{n}{2})}({\bf k}) = \begin{pmatrix}
		\alpha k_-^{\frac{n}{2}} & \beta k_+^{\frac{n}{2}-1}\\
		(-1)^{\frac{n}{2}} \beta k_-^{\frac{n}{2}-1} & (-1)^{\frac{n}{2}+1} \alpha k_+^{\frac{n}{2}}  \\
	\end{pmatrix},\ \ \ 
	h_c^{(0,0)}({\bf k}) = \begin{pmatrix}
		A & B k_- \\
		B k_+ & -A \\
	\end{pmatrix},
\end{eqnarray}
where we have assumed $\alpha,\beta,A,B$ to be real for simplicity.

\subsection{Edge theory of $H_\text{BL}({\bf k})$}

In this section, we write $h_{\pm}$ In the polar coordinates $(r,\theta)$ and further solve for its edge modes. Instead of imposing an open boundary condition at $r=R$, we can simplify the edge-mode problem to a geometry of chemical potential domain wall, by replacing $M({\bf k})$ with $\mu \text{sgn}(r-R)$ . Note that $k_{\pm} = e^{\pm i \theta} (k_r \pm i k_{\theta})$ with $k_r = -i \partial_r $ and $k_{\theta} = - \frac{i}{r} \partial_{\theta}$. When $R\gg a$ (a is the lattice constant), we further consider a small $k_{\theta}$ limit and treat all $k_{\theta}$-relevant terms as perturbations. Then the zeroth-order Hamiltonian is 
\begin{eqnarray}
	&& h_{s}^{(0)}(r,\theta) = \begin{pmatrix}
		s\mu \text{sgn}(r-R) & -i\Delta e^{-i\theta} \partial_r \\
		-i\Delta e^{i\theta} \partial_r &  -s\mu \text{sgn}(r-R) \\
	\end{pmatrix}, \nonumber \\
	&& h_{s}^{(1)}(r,\theta) = \begin{pmatrix}
	0 & -\Delta e^{-i\theta} \partial_r/r \\
	\Delta e^{i\theta} \partial_r/r &  0 \\
\end{pmatrix}
\end{eqnarray}
with $s=\pm$. An ansatz solution for $h_{s}^{(0)}$ takes a general form of
\begin{equation}
	|\varphi_l^{(s)}(r,\theta)\rangle = {\cal N} f(r-R) e^{il\theta} \chi_s (\theta),
\end{equation}
where the spatial part $f(r-R)$ describes a localized mode around $r=R$, $\chi_s(\theta)$ is a two-component spinor, and ${\cal N}$ is an overall normalization factor. $l\in\mathbb{Z}$ is the eigenvalue of $k_{\theta}$ and is related to the physical angular momentum of the edge mode. It is easy to show that 
\begin{equation}
	|\varphi_l^{(s)}(r,\theta)\rangle = {\cal N} e^{-|\frac{\mu}{\Delta}(r-R)|} e^{il\theta} \begin{bmatrix}
		e^{- i s \frac{\pi}{4}} \\
		e^{i(\theta + s \frac{\pi}{4})} 
	\end{bmatrix}.
\end{equation}
Crucially, the $z$-component angular momentum of $|\varphi_l^{(s)}(r,\theta)\rangle$ is 
\begin{equation}
	J_z = l+\frac{1}{2},
\end{equation}
which is a half integer. Projecting $h_s^{(1)}$ onto $|\varphi_l^{(s)}\rangle$ basis, we arrive at a pair of chiral edge modes with the dispersions,
\begin{equation}
	E_l^{(s)} = \frac{s \Delta}{R} (l+\frac{1}{2}),
	\label{eq-E_l}
\end{equation}
which admit a finite size gap $\sim \frac{\Delta}{R}$.

\subsection{Edge Gap and Corner MZMs}

The edge-state basis functions are defined as
\begin{equation}
	|\Phi_l^{(+)} \rangle = (|\varphi_l^{(+)}\rangle, 0)^T,\ \ |\Phi_l^{(-)} \rangle = (0, |\varphi_l^{(-)}\rangle)^T.
\end{equation}
Then the chiral edge mode Hamiltonian from Eq.~\ref{eq-E_l} is 
\begin{equation}
	h_\text{cem} = \frac{\Delta}{R} (l+\frac{1}{2}) \begin{pmatrix}
		1 & 0 \\
		0 & -1 \\
	\end{pmatrix}.
	\label{eq-cem}
\end{equation}
To visualize the existence of corner MZMs, we now project the interlayer coupling matrix $h_c^{(\nu,\nu')}({\bf k})$ onto $|\Phi_l^{(\pm)}\rangle$ basis, which leads to $h_g^{(\nu,\nu')}(\theta)$ that will explicitly gap out $h_\text{cem}$. For example, for $\Psi_\text{BL}=(\Psi_0, \Psi_0)^T$
\begin{eqnarray}
	h_g^{(0,0)}(\theta) = \begin{pmatrix}
		0 & i A + B \\
		-i A + B & 0 \\
	\end{pmatrix}, 
\end{eqnarray}
which clearly gaps out both edge modes. Therefore, both $\Psi_\text{BL}=(\Psi_0, \Psi_0)^T$ and $\Psi_\text{BL}=(\Psi_{\frac{n}{2}}, \Psi_\frac{n}{2})^T$ can only lead to a trivial superconductor phase, instead of a topological one. 

On the other hand, for $\Psi_\text{BL}=(\Psi_0, \Psi_\frac{n}{2})^T$, we find that the edge gap Hamiltonian has an interesting dependence on both $\theta$ and $n$. In particular,
\begin{equation}
	h_g^{(0,\frac{n}{2})}(\theta)  = \begin{cases}
		\begin{pmatrix}
			0 & (\alpha+i \beta) \sin \frac{n}{2} \theta \\
			(\alpha-i \beta) \sin \frac{n}{2} \theta  & 0 \\
		\end{pmatrix} \quad & n=2\ \&\ n=6 ,\\  \\
		\begin{pmatrix}
			0 &  (\beta + i \alpha) \cos \frac{n}{2} \theta \\
			(\beta - i \alpha) \cos \frac{n}{2} \theta & 0 \\
		\end{pmatrix} & n=4.
	\end{cases}
	\label{eq-hopping}
\end{equation}
Combing with Eq.~\ref{eq-cem}, we find that the edge gap closes whenever
\begin{equation}
	\theta = \begin{cases}
		\frac{2m\pi}{n} & \frac{n}{2} \equiv 1\ (\text{mod}\ 2), \\  \\ 
		\frac{(2m+1)\pi}{n} & \frac{n}{2} \equiv 0\ (\text{mod}\ 2),
	\end{cases}
\end{equation}
for $m=0,1,2,...,n-1$. This gap closing condition exactly corresponds to $n$ corner Majorana modes. Therefore, we conclude that 
\begin{itemize}
	\item Stacking a pair of chiral TSCs with opposite Chern numbers $|{\cal C}|=1$ will lead to an HOTSC$_n$, if the chiral TSCs feature Nambu bases $|\Psi_0\rangle$ and $|\Psi_{\frac{n}{2}}\rangle$, respectively. 
\end{itemize}

\end{document}